\documentclass[square]{ws-procs11x85}
\usepackage{multicol,float,balance} 

\begin{document}

\title{THE SPECTRAL SEQUENCE OF BLAZARS -- STATUS AND PERSPECTIVES }

\author{L. MARASCHI$^*$, G. GHISELLINI and F. TAVECCHIO}

\address{INAF/Osservatorio Astronomico di Brera, Via Brera 28, 20121, Milano, Italy\\
$^*$E-mail: laura.maraschi@brera.inaf.it}

\author{L. FOSCHINI}

\address{INAF/IASF--Bologna, 40129, Bologna, Italy}

\author{R. M. SAMBRUNA}

\address{NASA/Goddard Space Flight Center, Greenbelt, MD 20771, USA}

\begin{abstract}
The present status of the blazar spectral sequence is discussed, including new
findings about blazars selected with different criteria than the original 
complete radio-samples. Despite extensive searches of blazars "breaking" the
sequence, the original idea proposed 10 years ago, still seems to hold.
On the other hand the forthcoming launch of the \emph{GLAST} satellite will 
provide a new selection band for blazars and blazar related populations
as well as fantastic progress on the spectra and variability behaviour of 
presently known blazars. The order of magnitude increase in sensitivity of
\emph{GLAST} will allow to detect $\gamma$--rays from jets with lower power
and/or lower beaming factor, thus sampling a much wider population.
\end{abstract}


\bodymatter

\begin{multicols}{2}

\section{Introduction}\label{sec1}

The blazar spectral sequence was introduced 10 years ago \cite{fos98, ghi98}
in view of understanding the systematic differences in broad--band spectral 
properties from X--ray selected BL Lacs to radio--selected BL Lacs 
to Flat Spectrum Radio Quasars (FSRQ). 
The spectral energy distributions (SEDs)
of all these sources (unified under the term ``blazars") are dominated by the
 beamed non-thermal emission of a relativistic jet.  
From the results of the {\it Compton Gamma--Ray Observatory}, the $Beppo$SAX
satellite and Cherenkov telescopes, it emerged that all SEDs 
were characterized by two peaks, 
commonly attributed to Synchrotron and inverse Compton (IC) radiation 
respectively, emitted by a population of relativistic electrons,
with seed photons possibly provided by the synchrotron radiation 
itself (SSC), the accretion disk or the broad line region (EC).
Both peaks appeared to shift to lower frequency from the first class 
to the next. 

Phenomenologically, all the sources of the three complete samples
available at the time were grouped together and divided in radio luminosity
bins (each spanning a decade) irrespective of the original 
classification and the average SEDs for each bin were computed.
The results strongly suggested that the source's luminosity is a fundamental
parameter driving the overall shape of the SEDs, hence the concept of a
``blazar spectral sequence", implying a link of the main spectral and evolutionary 
properties of blazars with the physics of jets and the ``central engine".

Modelling the SEDs of individual objects indicated that the physical
parameters of the radiating region in the jets vary systematically
with {\it increasing} luminosity in the sense of an {\it increase} of the 
radiation energy density of the seed photons available for the inverse Compton process
and a {\it decrease} of the energy of the electrons contributing most
to the emission. 
A complete description of the spectral modelling used here
is given in \cite{ghi02, CG08}. 
The model includes synchrotron and Inverse Compton radiation
from a population of relativistic electrons whose energy distribution is 
determined by injection and energy losses in a finite time $R/c$ where $R$ is the
size of the emitting region. 
Jets immersed in strong radiation fields will then
have electron distributions with spectral breaks at lower energies.     
\begin{figure*}
\vskip -1 cm
\centerline{\psfig{file=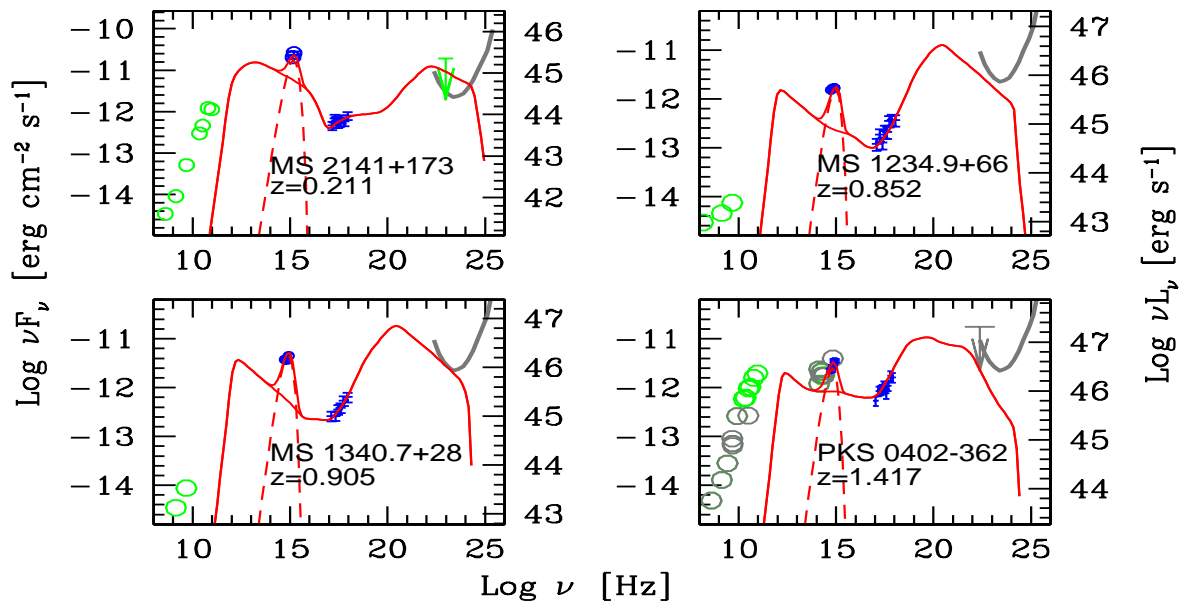,width=17cm,height=14cm}}
\vskip -5 cm
\caption{SEDs of the 4 EMSS blazars discussed in this paper, together with the
results of our modelling. The dashed line is the assumed contribution of the accretion
disk emission.
The grey stripe is the sensitivity (5$\sigma$) of {\it GLAST} for one year of 
exposure time.}
\label{emss}
\end{figure*}

The blazar sequence can also be understood in terms of a cosmological 
decrease in the average accretion rate $\dot{m}$ onto the central SMBHs.
At high--redshift, FSRQs are characterized by high accretion rates 
and $L_{\rm jet} \simeq L_{\rm disk}$; 
on the other hand, at low--redshift, BL Lac objects do not show signatures of
an accretion disk in the optical band and $L_{\rm jet} > L_{\rm disk}$, 
implying low radiative efficiency for the accretion flow (i.e. largely 
sub--Eddington accretion rates) unless one is prepared to accept that the
jet power {\it largely exceeds} the accretion power. 
The transition from
near--Eddington to sub--Eddington accretion rates with decreasing redshift is
naturally explained by the cosmological decrease in the availability of gas 
and increase in the black hole masses at the centers of galaxies 
\cite{mar01, bot02, cav02, mar03, sam06a}. 
It is important to note that the same 
conceptual trend could apply to the FRI/FRII dicothomy \cite{ghi01}.

%

It is therefore of paramount importance to test and extend the original 
blazar sequence suggestion, by removing the bias introduced by the incomplete 
and non--homogeneous coverage at $\gamma$--ray energies and by producing 
independent and larger samples of blazars. With the advent of \emph{GLAST}
both aspects will receive unprecedented momentum.

Here we discuss new objects with peculiar characteristics, found as a result
of systematic programs or serendipitously, which  may already challenge 
the sequence concept.

\section{New data, new sources}\label{sec3}


One key--prediction of the sequence concept is the {\it lack of FSRQ with a 
synchrotron peak in the UV/X--ray energy band}; therefore such objects have
been extensively searched (see \cite{pad07} and references therein).
Some authors\cite{pad02, bas07, gio07} claimed the discovery of such
``anomalous" blazars mainly on the basis of their broad band spectral 
indices, that is ratios of radio/optical/X--ray fluxes. 
However a knowledge of the spectra {\it within} each band is essential 
to confirm the claim. 

For the above reasons we recently started new observations with 
\emph{Swift} of FSRQs in the only {\it X--ray selected} sample of broad 
line radio loud AGN \cite{wol01}, derived from the \emph{Einstein 
Medium Sensitivity Survey} (EMSS)\cite{gio90}. 
X--ray selection allows to probe Radio Loud
Quasars 10--100 times weaker in the radio than classical samples and by
analogy with the case of X--ray selected BL Lacs, Radio Loud Quasars with a
high frequency synchrotron peak could be expected. 
Moreover, \emph{Swift}
provides not only spectral data in the X--ray band but also simultaneous 
fluxes in different optical/UV  bands, yielding precious information on the
shape of the optical to X--ray SED.

\subsection{EMSS Blazars}





The optical and X--ray data obtained  for the first 4 FSRQs 
observed with \emph{Swift}
are presented in Fig. \ref{emss} together with archival radio data and theoretical 
models for the SEDs.
The objects are ordered in redshift and present very similar SEDs:
in all cases the optical emission is likely associated with the ``blue bump", 
that is the thermal emission from the accretion disk.
The X--ray emission is rather hard suggesting an IC component with ``external"
seed photons if ascribed to the jet, as chosen in Fig 1.
However a contribution of X-ray emission from the accretion disk is possible. 
The strength of the non thermal (jet) emission relative to
the accretion disk, as estimated in Fig 1, increases with increasing redshift. 
This may suggest that at low redshifts, where lower 
X--ray luminosities can be detected, sources with weaker jets relative 
to the intensity of the accretion disk are found. 
The weakness may be intrinsic or due to a larger
viewing angle causing a lower beaming factor.

The $\gamma$--ray fluxes predicted from the models are uncertain, as they depend
on the shape of the relativistic electron spectrum at relatively modest
energies, where the constraints are poor, as well as on the photon density at 
the site of the emission region. 
There is thus some freedom in the parameter choice and we have followed the
criterion to minimize the observed bolometric luminosity.
Despite this ``economic" choice, the predicted $\gamma$--ray emission 
falls close to the \emph{GLAST} sensitivity limit, which is also shown in Fig. 1.  
Taking into account variability, it is likely that some of
these objects will be detected by \emph{GLAST}, in particular MS~$0402-362$
which has the highest X--ray to optical ratio, i.e. the largest jet to 
accretion power ratio. 
The \emph{GLAST} data would provide important constraints to the model parameters.

\subsection{The role of the viewing angle and GLAST }

\begin{figure}[H]
\vskip -0.5 cm
 \centerline{\psfig{file=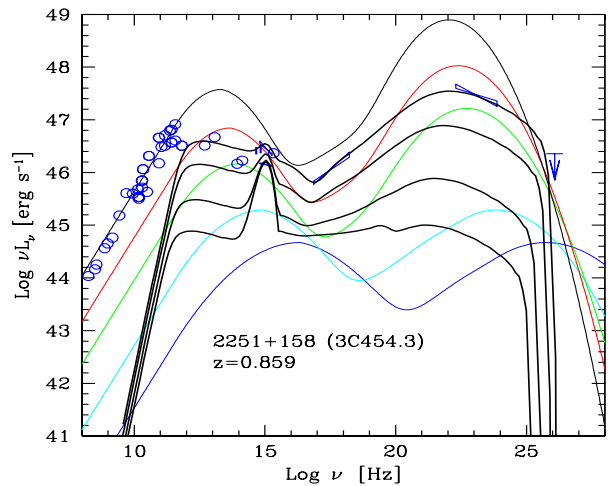,width=9.5cm,height=8cm}}
\caption{Example of how the SED of a blazar (3C 454.3) changes by
changing the viewing angle, hence the Doppler factor.  Note that the
EC component varies the most, since this emission is anisotropic even
in the comoving frame.  We show for comparison the SED corresponding of
the blazar sequence.}
\label{theta}
\end{figure}

In order to estimate the effect of different viewing angles on the appearence
of the SED of a given object with respect to the blazar sequence we use 
the SED model for the well known blazar 3C 454.3 \cite{tavjet07}. 
We compute model SEDs
with the same physical parameters but different viewing angles ($<10^\circ$), 
assuming a homogeneous jet. 
The results are shown in Fig. \ref{theta} overlayed on  the sequence SEDs. 
Clearly the $\gamma$--ray emission depends strongly on
the viewing angle while the synchrotron emission is less affected and
the blue bump luminosity remains constant. We conclude that the very 
existence (up to now) of the sequence is probably due to the fact that the 
sensitivity limits of the initial surveys only allowed the most beamed 
objects to be detected. This situation will change dramatically with 
\emph{GLAST} which will be able to detect blazars
at larger viewing angles. 
We also recall that ``structured jets" \cite{gtc05} may
show significant $\gamma$--ray emission over a larger range of angles.

\subsection{New very luminous blazars} 

A number of highly luminous (high $z$) blazars have been discovered
recently (e.g.  GB 1428+4217, $z = 4.72$ \cite{fabian98}; PMN
J0525--3343, $z=4.4$ \cite{fabian01}; RX J1028.6--0844, $z=4.276$
\cite{yuan00}; Q0906+6930, $z=5.47$ \cite{romani04}; RBS 315, $z=2.69$
\cite{tav07}). All exhibit a very hard X-ray spectrum, which
indicates an inverse Compton origin for the high energy radiation in
agreement with the sequence trends. However few data at hard X-rays
exist for these objects. Instead, the {\it BAT} instrument
onboard {\it Swift}, discovered a new "hard X-ray selected" blazar,
SDSS J074625.87+244901.2, with extremely high luminosity. Its spectrum in the
10-100 keV band points to a largely dominant inverse Compton peak in agreement
with expectations from the sequence \cite{maraglast}.

Two extremely high luminosity objects discovered recently: 
SDSS J081009.94+384757.0 ($z=3.945990$; \cite{gio07}) and 
MG3 J225155+2217 ($z=3.668$; \cite{bas07}) 
have been claimed to show a {\it synchrotron} peak in hard X--rays,
violating the sequence dramatically. 
If correct, this suggestion would require extremely high values of both,
magnetic field and particle energies. 

\begin{figure}[H]
\vskip -0.5 cm
\centerline{\psfig{file=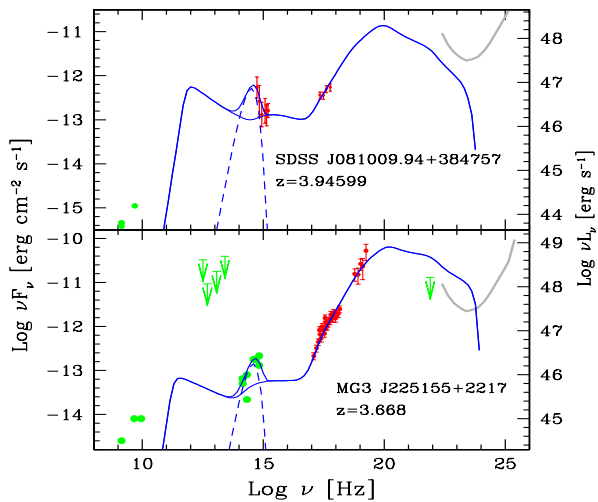,width=9.5cm,height=8cm}}
\vskip -0.8 cm
\caption{The SEDs of the two high--$z$ blazars discussed in the text,
together with the proposed models.}
\label{red}
\end{figure}

In order to verify these claims, we reanalised the existing data from
\emph{Swift} and \emph{INTEGRAL} to better constrain the SEDs of these
two sources.  The results are displayed in Fig \ref{red}.  In both
cases the SED from the optical to the X--ray range appears to be
concave pointing to an Inverse Compton origin for the X--ray to
$\gamma$--ray emission as in the ``standard'' model for the sequence
SEDs \cite{ghi98}.  While hard X-ray data for SDSS J081009.94+384757.0
are lacking, the \emph{INTEGRAL} data for MG3 J225155+2217, show that
its SED compares well with SDSS J074625.87+244901.2. Both objects are
extreme in luminosity and hard X--ray to optical ratio, suggesting an
extension of the sequence to higher luminosities in the sense of an
even higher Compton ``dominance" and a $\gamma$--ray peak at
relatively low energies, in the 1--10 MeV range.  \emph{GLAST}
observations will be crucial to understand the high energy emission
of these sources.  Note that, at the intensity level recorded recently
a \emph{GLAST} detection of J225155+2217 is expected  despite
its soft spectrum in the 10 MeV--10 GeV range, while the same is not true
for J081009.94+384757.0.

\subsection{Intermediate luminosity blazars peaking at high frequencies}

\begin{figure}[H]
\vskip -0.5 cm
\centerline{\psfig{file=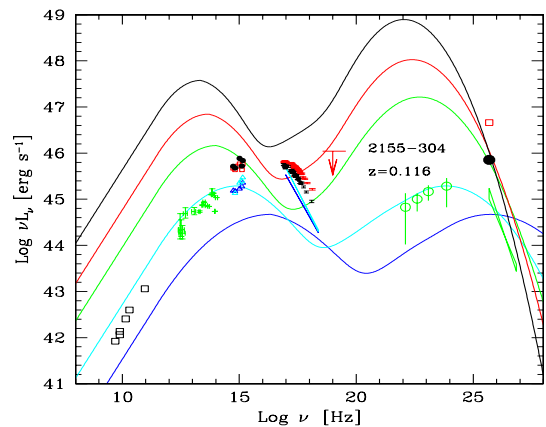,width=8.5cm,height=7cm}}
\caption{The SED of PKS 2155--304 (points) compared to the the average
SEDs corresponding to the blazar sequence. See text for details.}
\label{2155}
\end{figure}

\begin{figure}[H]
\vskip -0.5 cm
\centerline{\psfig{file=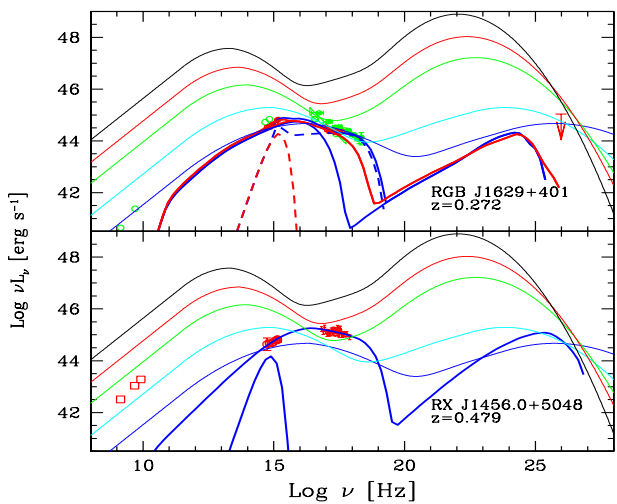,width=9.5cm,height=8cm}}
\vskip -0.8 cm
\caption{The SED of the two ``blue" quasars discussed in the text,
together with the proposed model.  For RGB J1629+401 we show two
possible models, to show (blue) the effect of a possible X--ray corona
associated to the disk accretion luminosity.  In both cases we have
superposed (thin solid lines) the average SEDs corresponding to the
blazar sequence.  }
\label{blue}
\end{figure}

Let us discuss first the well known BL Lac object PKS 2155--304 (see
Fig.\ref{2155}). This object is similar to, but more luminous than Mkn
421 and Mkn 501. In fact its ``normal'' state as described by the data
of \cite{fos07} (green dots) falls well on the sequence with the
synchrotron peak around $10^{16}$ Hz \cite{maraglast}, while on
average Mkn 421 and Mkn 501 peak at somewhat higher frequencies, in
agreement with the sequence. However, during flares, all these objects
exhibit a trend of {\it increasing peak frequency with increasing
luminosity,  contrary to the sequence trends}. Thus we caution that
the sequence ``concept'' applies only to average states.  In the
exceptional flaring state of July 28 2006, PKS 2155--304 changed its
peak intensity by almost one order of magnitude, thus moving from
luminosity class 2 to 3 and breaking the sequence during its high
state.
The source RX J1456.0+5048 ($z=0.478567$) discovered by Giommi et al.
(2008 in preparation) is close in properties to PKS 2155--304.  The
\emph{Swift} data analysed by us are shown in Fig. \ref{blue}
superimposed on the sequence average SEDs.  The object exhibits a peak
at high frequency ($10^{16}$ Hz) with a rather large peak X--ray
luminosity ($10^{45}$ erg s$^{-1}$), falling in the central region of
the sequence where the average SEDs nearly cross each other.  The
optical spectrum of this source exhibits very weak emission lines,
approaching a BL Lac classification which reinforces the similarity to
PKS 2155--304.

\subsection{The intriguing case of RGB J1629+401}

This blazar is particularly interesting, since it exhibits very
unusual properties. It is known  since some years and was
suggested to be a FSRQ ($z=0.271946$) with a synchrotron peak at
X--ray energies \cite{pad02}. In fact its SED, shown in Fig. 4 falls
in the same region as PKS 2155--304 and RX J1456+504.

\begin{figure}[H]
\vskip -1.8 cm
\hskip -0.2  true cm
\centerline{\psfig{file=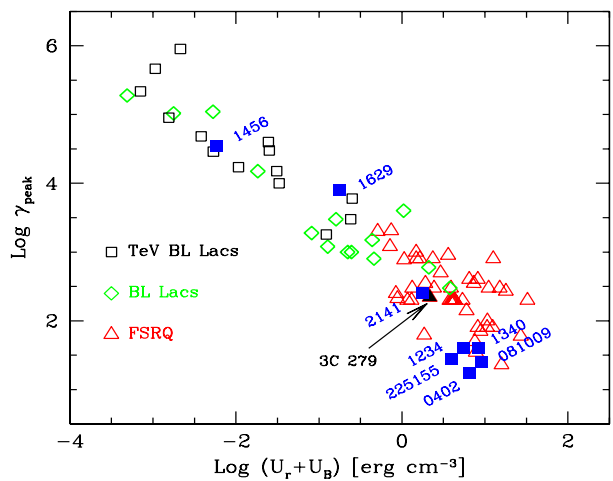,width=9.8cm,height=9cm}}
\vskip -0.8 cm
\caption{ The relation between $\gamma_{\rm peak}$ and the total
(magnetic plus radiative) energy density as seen in the comoving
frame.  We have labelled the sources discussed here and compare them
to the sample of blazars discussed in \cite{CG08}.  }
\label{gammau}
\end{figure}

Its optical spectrum shows strong but narrow emission
lines\footnote{\texttt{http://cas.sdss.org/astrodr6/en/tools/explore/obj.asp?ra=247.2554958\&dec=40.1332306}}.  Due to its high
optical luminosity ($M=-23$) it was classified as a radio--loud Narrow Line
Type 1 Quasar by Komossa et al.\cite{kom06, zho02} in a systematic
search for radio--loud NL objects of which 
NL Seyfert 1 galaxies are the most common.  The mass of the black
hole is estimated to be $2\times 10^7$ solar masses and  its
X--ray luminosity is highly super--Eddington.
However if the X--ray luminosity is attributed to a beamed jet, as
suggested by the SED and by the strongly ``inverted" spectral index in
the radio, the latter estimate may be reduced.  We note that the
radiation energy density $U_{\rm ext}$ derived from the observed
luminosity of the narrow emission lines, assuming a distance from the
black hole consistent with the line width ($v <2000$ km/s) yields a
value lower than found in broad line quasars.  Since $U_{\rm ext}$
is small, our model associates to this blazar a relatively large
$\gamma_{\rm peak}$, and then this source falls in the middle of the
$\gamma_{\rm peak}\propto U^{-1}$ branch of Fig. \ref{gammau}, though
technically it can be defined as a quasar from the point of view of the
brightness of its optical nucleus.  We therefore believe that
it is reasonable to hypothesize that this object hosts a strong
relativistic jet.  We recall that NL Type 1 quasars are rare among
quasars and "radio loudness" is rarer in NL quasars than in BL
quasars \cite{kom06}. Thus it is not surprising that such objects were not 
found in the radio--loud samples explored previously. Clearly RGB J1629+401
poses a number of questions that will need further studies.  A
detection at $\gamma$-ray energies could confirm the jet model but is not
guaranteed at this average intensity level.


\section{Discussion and Conclusions}

Despite the efforts to search for objects which may violate the sequence
trends no ``strong outliers" have been found. Some claims from other
authors seem unjustified, as we have shown.

The existence of the sequence can be traced back to a strong physical link
between the energy of the electrons emitting most of the power and the
total (radiative plus magnetic) energy densities. As discussed by
\cite{ghi98} and \cite{ghi02} this trend can be the result of the balance
between the cooling rate (measured by the amount of total energy
density) and the (almost universal) acceleration rate of the
electrons.  The most powerful sources have a large amount of magnetic
and radiation energy density, determining a severe cooling and thus a
small value for the equilibrium Lorentz factor of the electrons. On
the contrary, BL Lacs are characterized by a low level of cooling,
explaining the large electron Lorentz factors in these
sources. 

In Fig. \ref{gammau} the parameters derived for the sources discussed 
in this paper are compared with those determined for the group of blazars recently
considered by \cite{CG08}.
Although  RGB J1629+401 does not deviate significantly from other blazars in this
parameter plane it should be stressed that its other characteristics, notably
the estimated black hole mass and optical narrow line spectrum are at variance 
with previously known blazars.
Searching for more objects like RGB J1629+401, could extend the 
``blazar phenomenon'' to low masses and provide useful hints to understand
the link between relativistic jets and black hole masses.

High--$z$ objects show extreme ratios of inverse-Compton to
synchrotron emission, which -- with reference to the adopted SED
model -- corresponds to a high energy density of external radiation.
It is easy to anticipate that in the study of this external Compton component 
 \emph{GLAST} will play a crucial role.

\section*{Acknowledgments}

This work has made use of the \emph{NASA/IPAC Extragalactic Database
(NED)}, which is operated by the \emph{Jet Propulsion Laboratory},
\emph{California Institute of Technology}, under contract with the
\emph{National Aeronautics and Space Administration}, and of data
obtained from the \emph{High Energy Astrophysics Science Archive
Research Center (HEASARC)}, provided by \emph{NASA's Goddard Space
Flight Center}.  This work has been partially supported by contracts
ASI-INAF I/023/05/0 and I/088/06/0.

\end{multicols}
\end{document}